\documentclass[pra, twocolumn, superscriptaddress, reprint]{revtex4-1}

\usepackage{amsmath} 
\usepackage{amsfonts}
\usepackage{amssymb}
\usepackage{graphicx}  
\usepackage{hyperref}  
\usepackage{color}

\usepackage[table]{xcolor}
\usepackage{array,ragged2e}

\begin{document}
\bibliographystyle{apsrev4-1}

\title{Position, spin and orbital angular momentum of a relativistic electron}

\author{Konstantin Y.~Bliokh}
\affiliation{Center for Emergent Matter Science, RIKEN, Wako-shi, Saitama 351-0198, Japan}
\affiliation{Nonlinear Physics Centre, RSPE, The Australian National University, Canberra, Australia }

\author{Mark R.~Dennis}
\affiliation{H.~H.~Wills Physics Laboratory, University of Bristol, Bristol BS8 1TL, UK}

\author{Franco Nori}
\affiliation{Center for Emergent Matter Science, RIKEN, Wako-shi, Saitama 351-0198, Japan} 
\affiliation{Physics Department, University of Michigan, Ann Arbor, MI 48109-1040, USA}

\begin{abstract}
Motivated by recent interest in relativistic electron vortex states, we revisit the spin and orbital angular momentum properties of Dirac electrons. 
These are uniquely determined by the choice of the {\it position operator} for a relativistic electron. 
We consider two main approaches discussed in the literature: (i) the projection of operators onto the positive-energy subspace, which removes the {\it zitterbewegung} effects and correctly describes spin-orbit interaction effects, and (ii) the use of Newton-Wigner-Foldy-Wouthuysen operators based on the inverse Foldy-Wouthuysen transformation. 
We argue that the first approach [previously described in application to Dirac vortex beams in K.Y.~Bliokh {\it et al}., Phys.~Rev.~Lett.~107, 174802 (2011)] has a more natural physical interpretation, including spin-orbit interactions and a nonsingular zero-mass limit, than the second one [S.M.~Barnett, Phys.~Rev.~Lett.~118, 114802 (2017)].
\end{abstract}

\maketitle

\section{Introduction}
\label{sec:1}

Impressive achievements in investigations and applications of optical vortex beams and optical angular momentum (AM) \cite{1,2} motivated the prediction and generation of free-electron vortex states carrying orbital angular momentum \cite{3,4,5,6}. 
These electron vortices are currently attracting considerable attention in several areas of physics, including electron microscopy, quantum theory, and high-energy physics (see \cite{7,8} for reviews). 
Free-electron vortex states were first described using a simplified model of a scalar non-relativistic electron \cite{3}. 
Soon after the generation of electron vortex beams in transmission electron microscopes \cite{4,5,6}, we provided a fully relativistic description of vortex electrons with spin by constructing exact Bessel-beam solutions of the Dirac equation (Dirac-Bessel beams) \cite{9}. 
Using a covariant position operator for the Dirac electron, we also introduced the corresponding separately-conserved spin and orbital AM of a relativistic electron, and described observable spin-orbit interaction (SOI) effects. 
Later, Dirac-Bessel electron beams were employed in the contexts of high-energy physics, scattering, and radiation problems \cite{8,10,11,12,13,14,15}.

In fact, the scalar-model description is still the most suitable for electron-microscopy applications. 
First, electron beams in TEMs are \emph{unpolarized}. 
Second, these are strongly \emph{paraxial}, and SOI effects become negligible under such conditions \cite{16}. 
Nevertheless, there is still theoretical interest in relativistic electron vortex states, and two recent works \cite{17,18} revisited vortex solutions and AM properties of the Dirac equation, in slightly different contexts to \cite{9}.

First, Bialynicki-Birula and Bialynicka-Birula \cite{17} suggested an elegant way to construct Dirac-Bessel beams \cite{9} and other vortex solutions of the Dirac equation. 
In particular, the authors introduced vortex ``wave packet'' solutions, which are actually localized only in \emph{time} but unbounded along the longitudinal $z$ coordinate. Therefore, such solutions cannot model longitudinally-localized electron wave packets in typical experimental conditions, but they could be useful in problems where the finite width of the electron energy spectrum is crucial. 
Note that usually electron beams in electron microscopes are modeled via an integral of Bessel beams over different transverse \emph{momenta} \cite{8,19} rather than \emph{energies} \cite{17}. 
In any case, Dirac-Bessel beams still represent the main building blocks for relativistic electrons, and superpositions of such states with different momenta or/and energies provide properly localized solutions. 
The authors of \cite{17} also emphasized the non-singular character of the vorticity based on the Dirac probability current, i.e., the absence of a well-defined vortex core in Dirac vortex beams. 
Here we should note that Dirac vortex beams represent \emph{vector} beams (or, more accurately, spinor beams), which should be characterized via \emph{polarization singularities} \cite{20} rather than the simple vorticity used for scalar wave fields, and there is no natural analogue to the polarization singularities of spin-1 waves like light for spin 1/2 particles. 
Nonetheless, such vector beams clearly exhibit vortices in each component of the spinor wavefunction and they carry well-defined (but non-integer) orbital AM \cite{9}. 
Finally, the Dirac current is naturally decomposed into the orbital and spin parts \cite{17} via the Gordon decomposition, although we emphasize a crucial difference with the analogous decomposition for optical fields (photons): while the orbital (canonical) current is observable in monochromatic optical fields (because it is directly coupled to dipole particles or atoms) \cite{21,22,23}, only the total Dirac (kinetic) current is observable in experiments with electrons (because it corresponds to the electric current coupled to electromagnetic fields).

Second, Barnett \cite{18} used approximate \emph{paraxial} Laguerre-Gaussian beam solutions of the Dirac equation to characterize relativistic electron vortex states. 
In the regime considered, the transverse momenta are assumed to be negligible compared to the mass and longitudinal momentum. 
However, SOI effects appear in non-paraxial corrections to the paraxial regime \cite{9}, and their accurate analysis requires full non-paraxial solutions, just as for solutions of Maxwell's equations \cite{24,25}. 
In this manner, a typical non-paraxial vortex solution of the Dirac equation with vortex charge $\ell$ in the main component (surviving in the paraxial limit) acquires extra components with vortices of charge $\ell + 2s_z$, where $s_z = \pm 1/2$ corresponds to the two states of the longitudinal projection of the rest-frame spin of the electron \cite{8,9,10,11,12,26,27}. 
However, only one of these components is present in \cite{18}, and, furthermore, calculations of the expectation value of the operator $(\mathbf{r} \times \boldsymbol{\alpha})_z$ (where $\boldsymbol{\alpha}$ is the usual matrix operator characterizing the Dirac probability current) yielded $\ell / E$ (where $E$ is the electron energy). 
However, this quantity describes the $z$-component of the \emph{magnetic moment} of the electron \cite{28,29}, and in the paraxial regime in \cite{9} it was found to be $(\ell + 2 s_z)/ E$, where $s_z$ is the expectation value of the longitudinal spin component. 
This latter expression correctly incorporates the expected g-factor of 2 for the electron spin.

Most importantly, alternative separately-conserved spin and orbital AM operators were suggested in Ref.~\cite{18}, which do not exhibit any signature of SOI and yield expectation values different from those obtained in Ref.~\cite{9}. 
The nontrivial differences between these treatments, and particularly the two apparently conflicting descriptions of the spin and orbital AM of the Dirac electron in Refs.~\cite{9} and \cite{18} motivated the present study. 
Here we show that the choice of the spin and orbital AM operators for the Dirac electron is uniquely determined by the choice of the \emph{position operator} for a relativistic electron. 
This is a long-standing problem analyzed in detail in a number of earlier works \cite{30,31,32,33,34,35,36,37,38,39,40,41,42}, starting with the seminal paper \cite{30} by Pryce in 1948. 
It is not surprising that the position operator is not completely straightforward in relativistic quantum mechanics: as position does not commute with the Dirac Hamiltonian, its action mixes positive and negative energy Fourier components of the Dirac equation, giving rise to \emph{zitterbewegung}. 
In short, there are two main approaches, each of which was used in \cite{9} and \cite{18}, respectively:

(i) The operators under discussion, including position, spin, and orbital AM, are \emph{projected} onto the direct sum of positive-energy (electron) and negative-energy (positron) subspaces. 
This does not change observable expectation values for pure electron states but ``corrects'' the time evolution of observables by removing \emph{zitterbewegung} effects \cite{33,34,35,36,37,38}. 
Such an approach results in the Berry-phase formalism commonly used for the description of various SOI phenomena for both relativistic spinning particles, including photons, and quasiparticles in solids \cite{24,25,39,40,41,43,44,45}.

(ii) Alternative operators are obtained via the ``\emph{Newton-Wigner-Foldy-Wouthuysen}'' (NWFW) approach, which is based on the inverse Foldy-Wouthusen (FW) transformation of the canonical operators \cite{31,32,34,35,36,37,42}. 
The Dirac Hamiltonian is diagonal in the FW representation, and so it might seem natural to define the position and other operators to have canonical forms in this representation, giving rise to the NWFW operators introduced in \cite{31,32}. 
It is worth noting, however, that the FW representation is problematic in full quantum electrodynamics, where the electromagnetic field is minimally-coupled to the electron characteristics in the standard (Dirac) representation.

In this work, we present an overview of various position and AM operators in different representations, emphasizing their properties, physical meaning, and observable manifestations. 
We argue that the NWFW approach has drawbacks compared to the ``projection'' formalism. 
Namely, it changes the observable expectation values of the quantities and also has a singular massless limit $m \to 0$, i.e., cannot be used for massless particles (e.g., photons). 
We show that this approach is essentially related to the \emph{rest-frame} characteristics of the electron, and therefore lacks the observable relativistic SOI phenomena. 
Throughout the paper, our treatment is based on the first-quantized (wave) approach to the Dirac equation.

\section{Basic equations}\label{sec:2}

We work with the standard representation of the Dirac equation in units with $\hbar = c = 1$: 
\begin{equation}
   i \frac{\partial \psi}{\partial t} = H \psi, \qquad H = \boldsymbol{\alpha}\cdot\mathbf{p} + \beta m,
   \label{eq:1}
\end{equation}
where $\psi(\mathbf{r},t)$ is the bispinor wavefunction, $H$ is the Dirac Hamiltonian, and
\[  \boldsymbol{\alpha} = \left(\begin{array}{cc} 0 & \boldsymbol{\sigma} \\ \boldsymbol{\sigma} & 0 \end{array}\right), \qquad
\beta = \left(\begin{array}{cc} 1 & 0 \\ 0 & -1 \end{array}\right)  \]
are the $4\times 4$ Dirac matrices \cite{37,46}. 
The four-component Dirac wavefunction implies four independent bispinor ``polarizations'', which correspond to two spin states with \emph{positive} energy (describing electrons) and two spin states with \emph{negative} energies (corresponding to positrons in the first-quantization approach we use).

The positive-energy (electron) plane-wave solutions of Eq.~(\ref{eq:1}) are $\psi_{\bf p}^{e} = W({\bf p}) \exp(i {\bf p}\cdot {\bf r} - i E t )$, with the bispinor 
\begin{equation}
   W = \frac{1}{\sqrt{2E}} \left( \begin{array}{c} \sqrt{E+m}\, w \\ \sqrt{E-m}\, \boldsymbol{\sigma}\cdot\bar{\bf p}\, w \end{array}\right).
   \label{eq:2}
\end{equation}
Here, $E = \sqrt{m^2 +p^2} >0$, $\boldsymbol{\sigma}$ is the 3-vector of Pauli matrices, $\bar{\bf p} = {\bf p}/p$ is the unit vector along the momentum direction, and $w = (a,b)^T$ is the two-component polarization spinor, $w^{\dagger}w = 1$, describing the spin state of the electron \cite{37,46}. 
The fact that the bispinor (\ref{eq:2}) has non-zero lower components for $\mathbf{p} \neq 0$ means that the standard representation is not diagonal with respect to the electron and positron subspaces, and describing pure \emph{electron} properties requires some care.

The positive- and negative-energy subspaces can be separated in the momentum representation using the unitary Foldy-Wouthuysen (FW) transformation which diagonalizes the Dirac Hamiltonian \cite{32,37,46}:
\begin{eqnarray}
   \psi_{\mathrm{FW}} & = & U_{\mathrm{FW}}(\mathbf{p})\psi, \quad H_{\mathrm{FW}} = U_{\mathrm{FW}} H U_{\mathrm{FW}}^{\dagger} =\beta E, \nonumber \\ 
   U_{\mathrm{FW}} & = & \frac{E+m+\beta \boldsymbol{\alpha}\cdot\mathbf{p}}{\sqrt{2E(E+m)}}. 
   \label{eq:3}
\end{eqnarray}
The bispinor (\ref{eq:2}) of the electron (positive-energy) plane wave has only upper components in this representation: $W_{\mathrm{FW}} \propto (w,0)^T$. 
Although the FW transformation is momentum-dependent and hence is nonlocal in real space, it is often convenient for the analysis of operators and calculations of their expectation values.

We focus on the angular momentum (AM) properties of a relativistic electron. 
The total AM operator $\mathbf{J}$ is well-defined for the Dirac equation:
\begin{equation}
   \mathbf{J} = \mathbf{r} \times \mathbf{p} + \mathbf{S} \equiv \mathbf{L} + \mathbf{S}, \qquad \mathbf{S} = \frac{1}{2}\left( \begin{array}{cc} \boldsymbol{\sigma} & 0 \\ 0 & \boldsymbol{\sigma} \end{array}\right),
   \label{eq:4}
\end{equation}
where $\mathbf{L}$ and $\mathbf{S}$ are canonical operators of the orbital and spin AM. 
It is well-known that the total AM $\mathbf{J}$ commutes with the Hamiltonian and thus is conserved, while $\mathbf{L}$ and $\mathbf{S}$ do not \cite{37,46}:
\begin{equation}
   [ H, \mathbf{J}] = 0, \quad [H,\mathbf{L}] \neq 0, \quad [H,\mathbf{S}] \neq 0. 
   \label{eq:5}
\end{equation}
This has led to considerable discussion and various suggestions on how to describe the spin and orbital AM of the Dirac electron.

As first realized by Pryce \cite{30}, since the operators $\mathbf{J}$ and $\mathbf{p}$ are uniquely defined and conserved for the free-space Dirac equation, the spin-orbital separation is intimately related to the choice of the \emph{position operator}. 
Indeed, choosing some position operator $\tilde{\mathbf{r}}$ determines the corresponding orbital AM $\tilde{\mathbf{L}} = \tilde{\mathbf{r}} \times \mathbf{p}$ and spin $\tilde{\mathbf{S}} = \mathbf{J} - \tilde{\mathbf{L}}$. 
Most significantly, the canonical position operator $\mathbf{r}$ is somewhat problematic for relativistic electrons as it corresponds to the velocity
\begin{equation}
   \frac{d \mathbf{r}}{d t} = i[H,\mathbf{r}]= \boldsymbol{\alpha}. 
   \label{eq:6}
\end{equation}
This velocity operator has eigenvalues $\pm 1$ and is in sharp contrast to the equation of motion of a \emph{classical} relativistic electron: $d\mathbf{r} / d t = \mathbf{p} / E$. 
This discrepancy is interpreted as \emph{zitterbewegung} oscillations produced by the interference in mixed electron-positron solutions \cite{37}. 
At the same time, the expectation value $\langle \mathbf{r} \rangle$ for a \emph{pure electron} (positive-energy) wavefunction $\psi^{e}({\bf r}, t)$ is meaningful
and does obey the proper equation of motion: $d \langle \mathbf{r} \rangle / dt = \langle {\bf p} H^{-1} \rangle$ \cite{37}. 
Similarly, the commutators (\ref{eq:5}) are non-zero due to \emph{zitterbewegung} effects \cite{37}, whereas the expectation values $\langle \mathbf{S}\rangle$ and $\langle \mathbf{L}\rangle$ for an electron wavefunction are meaningful observable quantities.

\section{Spin-orbit interaction}\label{sec:3}

Relativistic wave equations (including the Dirac and Maxwell equations) have inherent spin-orbit interaction (SOI) properties. 
In the Dirac equation, the SOI appears \emph{not} because of non-zero commutators (\ref{eq:5}), (i.e., not due to \emph{zitterbewegung}) since SOI is clearly manifest in the \emph{expectation values} of observable quantities for pure electron states. 
Below we show a few important examples. 

1. Consider the expectation value of the canonical spin operator $\mathbf{S}$, Eq.~(\ref{eq:4}), for an electron plane-wave state (\ref{eq:2}). 
For a motionless electron, $\mathbf{p} = 0$, and the bispinor $W$ has only the two upper components, given by the spinor $w$, which describes the \emph{non-relativistic} (in other words, rest-frame) electron spin: $\langle \mathbf{S} \rangle = W^{\dagger}\mathbf{S}W = w^{\dagger}\boldsymbol{\sigma}w / 2 \equiv \langle \mathbf{s} \rangle$. 
In particular, $w^+ = (1,0)^T$ and $w^- = (0,1)^T$ correspond to  $\langle s_z \rangle = \pm 1 / 2$, respectively. 
For a \emph{relativistic} electron with $\mathbf{p} \neq 0$, the expectation value of the spin becomes:
\begin{equation}
   \langle \mathbf{S} \rangle = \frac{m }{E}\langle \mathbf{s}\rangle + \frac{(\mathbf{p}\cdot\langle \mathbf{s}\rangle) \mathbf{p}}{E(E+m)}. 
   \label{eq:7}
\end{equation}
Up to the additional $m / E$ factor in the first summand, this equation coincides with a Lorentz boost of the spatial components of the four-vector $(0,\langle\mathbf{s}\rangle)$ from the electron rest frame to the laboratory frame \cite{46}. 
Thus, the expectation value of the relativistic electron spin is \emph{momentum-dependent}, which signals the SOI caused by relativistic transformations of the dynamical properties of the electron. 
This effect grows in significance in the ultrarelativistic (or massless) limit, when the momentum is large compared to the mass. 

2. Only momentum and spin can be determined for a single plane wave. 
Calculating other characteristics requires structured Dirac-electron solutions, such as \emph{electron vortex beams} \cite{3,4,5,6,7,8}. 
A monochromatic beam with well-defined energy $E > 0$, propagating along the $z$-axis, can be constructed as a Fourier superposition of multiple plane waves with momenta $\mathbf{p}$ distributed around the propagation direction:
\begin{equation}
   \psi({\bf r},t) \propto \int d^2\mathbf{p}_{\bot} W(\mathbf{p})f(\mathbf{p}_{\bot})e^{i\mathbf{p}\cdot {\bf r}-iEt}. 
   \label{eq:8}
\end{equation}
Here, $\mathbf{p}_{\bot} = (p_x, p_y)$ are the transverse momentum components describing deflections of the plane waves from the $z$-axis, while $f(\mathbf{p}_{\bot})$ is a scalar function describing Fourier amplitudes of the plane-wave components. 
For the simplest case of Bessel beams, the momenta are distributed on a circle in $\mathbf{p}$-space (lying on the appropriate mass shell),
\begin{equation}
   f_{\kappa,\ell}(\mathbf{p}_{\bot}) \propto \delta(p_{\bot} - \kappa) e^{i\ell \phi}, 
   \label{eq:9}
\end{equation}
where $(p_{\bot},\phi)$ are polar coordinates in the $\mathbf{p}_{\bot}$-plane, $\kappa$ determines the aperture angle $\theta_0$ of the beam ($\sin\theta_0 = \kappa / p < 1$), and $\ell$ is the integer azimuthal quantum number (vortex charge) of the beam. 
Although Bessel beams are not properly localized (square-integrable) with respect to the radial coordinate, they are very convenient for mathematical analysis. 
If radial localization is crucial, one can consider some appropriate Hankel integral of Bessel beams over $\kappa$, $\int_{\kappa_1}^{\kappa_2}  d\kappa$, whilst keeping the energy fixed (for $\kappa_1 \ll \kappa_2$ this corresponds to the physical electron vortex beams generated in transmission electron microscopes \cite{8,19}).

The polarization (spin) state of the beam (\ref{eq:8}) is specified by setting the spinor $w(\mathbf{p})$ for each plane wave in the spectrum. 
Assuming uniform polarization for all plane waves, i.e., $w$ independent of $\mathbf{p}$, our earlier calculations \cite{9} showed that the expectation value of spin and orbital AM in such electron vortex states become
\begin{equation}
   \langle S_z \rangle  =(1 - \Delta) \langle s_z \rangle, \quad \langle L_z \rangle = \ell + \Delta \langle s_z \rangle, 
   \label{eq:10}
\end{equation}
where $\Delta = (1 - m / E) \sin^2 \theta_0$ is the SOI parameter involving the beam aperture angle $\theta_0$. 
Equations (\ref{eq:10}) describe the SOI effect known as ``spin-to-orbital AM conversion'', which is well studied for non-paraxial light beams, both theoretically and experimentally \cite{24,25,47,48,49,50}. 
This shows that the orbital AM (and other observable orbital characteristics \cite{9,24,25}) of relativistic particles become \emph{spin-dependent}. 
Note that the expectation value $\langle S_z \rangle$ in Eq.~(\ref{eq:10}) follows from the plane-wave equation (\ref{eq:7}) after the substitutions $\langle {\bf s} \rangle = \langle s_z \rangle\, \bar{\bf z}$ and ${\bf p}\cdot\bar{\bf z} = p\, \cos\theta_0$, where $\bar{\bf z}$ is the unit vector along the $z$-axis. 
We also note that the SOI vanishes in the paraxial limit $\theta_0 \to 0$. 

3. Finally, the expectation value of the \emph{position operator}, $\langle \mathbf{r} \rangle$, can also reveal SOI effects
\cite{24,25,33,35,36,39,40,41,43,44,45}. 
First, the spin-Hall effect, well studied for electrons and photons in external potentials, appears as a semiclassical spin-dependent correction in the equation of motion for $d \langle \mathbf{r}\rangle / dt$ \cite{25,39,40,41,43,44,45,51,52,53}. 
Second, we explicitly calculate the expectation value $\langle \mathbf{r}_{\bot}\rangle$ for the transverse coordinates in the electron vortex beam. 
Obviously, $\langle \mathbf{r}_{\bot} \rangle = 0$ for any cylindrically-symmetric probability density distribution. 
For the same beam in a reference frame moving perpendicular to the beam axis with a relativistic velocity ${\bf v} \bot \bar{\bf z}$, the beam centroid drifts back as $-{\bf v} t'$, where primes denote quantities in the moving reference frame. 
Importantly, relativistic transformations of the angular-momentum tensor also require the centroid of the electron carrying intrinsic AM to be shifted in the direction orthogonal to both $\mathbf{v}$ and $\bar{\mathbf{z}}$ \cite{54,55}.  
As a result, the expectation value of the electron coordinate in the moving frame becomes \cite{54,55,56,57}
\begin{equation}
   \langle\mathbf{r}^{\prime}_{\bot}\rangle = -\mathbf{v} t' - \frac{\mathbf{v} \times \langle {\bf J} \rangle}{2E}, 
   \label{eq:11}
\end{equation}
where $\langle \mathbf{J} \rangle = (\ell + \langle s_z \rangle)\bar{\bf z}$ for the vortex beams considered above.
This AM-dependent tranverse shift induced by the transverse Lorentz boost can be called the \emph{relativistic Hall effect} \cite{55}, and it is closely related to the phenomena of Thomas precession, SOI, and AM conservation \cite{54}.

\section{Projected and ``Newton-Wigner- Foldy-Wouthuysen'' operators}\label{sec:4}

Having meaningful expectation values, one might wish to construct more meaningful position, spin, and orbital AM operators, free of zitterbewegung effects. 
As mentioned above, there are two main ways of doing this.

\subsection{Projected operators}\label{sub:4.1}

The most natural way to provide electron position, spin and orbital AM operators is to {\it project} these onto positive- and negative-energy subspaces, eliminating the cross-terms corresponding to the electron-positron transitions. 
This idea was first suggested by Schr\"odinger and can be written as \cite{36,37,38}
\begin{equation}
   \boldsymbol{\mathcal{R}} = \Pi^+\mathbf{r}\Pi^+ + \Pi^-\mathbf{r}\Pi^-, 
   \label{eq:12}
\end{equation}
where
\[ \Pi^{\pm} = \frac{1}{2} U_{\mathrm{FW}}^{\dagger} (1 \pm \beta) U_{\mathrm{FW}} = \frac{1}{2}\left(1 \pm \frac{m}{E} \beta\right) \pm \frac{\boldsymbol{\alpha}\cdot\mathbf{p}}{2E} \]
are the projectors onto the corresponding subspaces, and only the ``+'' subspace contributes to the expectation values for pure-electron states. Equation (\ref{eq:12}) yields the following projected position operator in the standard and FW representations \cite{9,36,37,38,39,40,41,43,44,45},
\begin{equation}
   \boldsymbol{\mathcal{R}} = \mathbf{r} + \frac{\mathbf{p}\times \mathbf{S}}{E^2} + i \frac{m \beta \boldsymbol{\alpha}}{2E^2}, \quad 
   \boldsymbol{\mathcal{R}}_{\mathrm{FW}} = \mathbf{r} + \frac{\mathbf{p}\times \mathbf{S}}{E(E+m)}. 
   \label{eq:13}
\end{equation}

The same projection procedure (\ref{eq:12}) can be applied to other operators. 
While it does not affect ${\bf p}$ and ${\bf J}$, the modified orbital and spin AM operators become
\begin{equation}
   \boldsymbol{\mathcal{L}}= \boldsymbol{\mathcal{R}} \times \mathbf{p}, \quad \boldsymbol{\mathcal{S}} = \mathbf{J} - \boldsymbol{\mathcal{L}}. 
   \label{eq:14}
\end{equation}
Explicitly, the projected spin operator is \cite{9,37,38,42}
\begin{align}
   \boldsymbol{\mathcal{S}} & = \frac{m^2}{E^2} \mathbf{S} + \frac{(\mathbf{p}\cdot\mathbf{S})\mathbf{p}}{E^2} - i \frac{m \beta(\boldsymbol{\alpha}\times\mathbf{p})}{2E^2}, \nonumber \\
   \boldsymbol{\mathcal{S}}_{\mathrm{FW}} & = \frac{m}{E} \mathbf{S} + \frac{(\mathbf{p}\cdot\mathbf{S})\mathbf{p}}{E(E+m)}. 
   \label{eq:15}
\end{align}
Importantly, the projected spin (\ref{eq:15}) corresponds to the spatial part of the \emph{Pauli-Lubanski 4-vector} $\mathcal{W}^{\mu}$, which correctly describes the spin states of moving relativistic particles (i.e., generating the little group of the Poincar\'e group) \cite{38,42}: 
\begin{equation}
   \boldsymbol{\mathcal{S}} = \boldsymbol{\mathcal W}H^{-1},~~ \mathcal{W}^{\mu} \equiv (\mathcal{W}^0 ,\boldsymbol{\mathcal W}) = \left(\mathbf{p}\cdot\mathbf{S},\tfrac{1}{2}(\mathbf{S}H+H\mathbf{S})\right).
   \label{eq:16}
\end{equation}
We also note that the operator $\boldsymbol{\mathcal{S}}_{\mathrm{FW}}$ explicitly reflects the structure of the expectation value of the relativistic electron spin $\langle \mathbf{S} \rangle$, Eq.~(\ref{eq:7}). 
Furthermore, since the electron wavefunction in the FW representation is reduced to the upper two components corresponding to the spinor $w$, the projected operators (\ref{eq:13})--(\ref{eq:15}) in the FW representation can be reduced to the $2 \times 2$ operators acting on the ``+'' subspace: $\mathbf{S} \to \mathbf{s} = \boldsymbol{\sigma}/2$.

The projected operators have two important properties. 
First, they obey \emph{proper time evolution} with conserved spin and orbital AM \cite{30,36,37},
\begin{equation}
   \frac{d \boldsymbol{\mathcal{R}}}{d t} = i [H,\boldsymbol{\mathcal{R}}]= \mathbf{p}H^{-1}, \quad [H,\boldsymbol{\mathcal{S}}]=[H,\boldsymbol{\mathcal{L}}] = 0. 
   \label{eq:17}
\end{equation}
Second, for any localized electron state they have the \emph{same expectation values} as the corresponding canonical operators,
\begin{equation}
   \langle \boldsymbol{\mathcal{R}} \rangle = \langle \mathbf{r} \rangle, 
   \quad \langle \boldsymbol{\mathcal{S}} \rangle = \langle \mathbf{S} \rangle, 
   \quad \langle \boldsymbol{\mathcal{L}} \rangle = \langle \mathbf{L} \rangle. 
   \label{eq:18}
\end{equation}
These follow automatically from the definition (\ref{eq:12}) assuming the states averaged over are pure electron states $\psi^{e}$ (i.e., already in the ``+'' subspace). 
Thus, \emph{the projection (\ref{eq:12}) affects the zitterbewegung phenomena for mixed electron-positron states but does not change observable quantities for pure electron states.}

Projection onto the ``$+$'' and ``$-$'' subspaces plays the role of a constraint, and it modifies commutation relations of the operators \cite{9,30,33,34,35,36,37,38,39,40} (cf.~\cite{24,58} for the photon analogs),
\begin{align}
   [\mathcal{R}_i,\mathcal{R}_j]  & =  - i \varepsilon_{ijk} \frac{S_k}{E^2},  \label{eq:19} \\
   [ \mathcal{S}_i, \mathcal{S}_j ]  & =  i \varepsilon_{ijk} \left( \mathcal S_k - \frac{(\mathbf{p}\cdot\boldsymbol{\mathcal{S}})p_k}{E^2}\right),\nonumber \\
   [\mathcal{L}_i,\mathcal{L}_j]  & =  i \varepsilon_{ijk} \left( \mathcal L_k - \frac{(\mathbf{p}\cdot\boldsymbol{\mathcal{S}})p_k}{E^2}\right).
   \label{eq:20}
\end{align}
In modern terms, one says that the projection generates a nontrivial {\it Berry connection} $\mathbf{A}_{B}(\mathbf{p})$ and
{\it curvature} $\mathbf{F}_{B} = -\boldsymbol{\mathcal S} / E^2$ in momentum space, resulting in covariant non-commutative coordinates (13) and (19): $\boldsymbol{\mathcal{R}} = \mathbf{r} + \mathbf{A}_{B}, [\mathcal{R}_i, \mathcal{R}_j ] = i \varepsilon_{ijk} F_{B k}$ \cite{9,24,39,40,41,43,44,45}. 

The above covariant (projected) operators underpin the modern theory of quasiparticles in solids (e.g., Bloch electrons) and relativistic spinning particles in external fields (including photons) \cite{9,24,39,40,41,43,44,45,51,52,53,59,60}. 
This approach has two great advantages. 
First, it describes \emph{observable Berry-phase and SOI phenomena}, from SOI Hamiltonians to Hall effects and topological states of matter. 
Second, the above approach \emph{can be equally applied to massive and massless particles}, i.e., it is not singular in the $m \to 0$ limit. 
In particular, for $m = 0$, Eqs.~(\ref{eq:13})--(\ref{eq:20}) become equivalent to the analogous equations for photons or classical light \cite{24,36,58,59,60}. 
The only difference is that in the photon case there is no negative-energy subspace, and the so-called \emph{transversality constraint} ($\mathbf{p}\cdot\mathbf{E} = \mathbf{p}\cdot\mathbf{H} = 0$, where $\mathbf{E}$ and $\mathbf{H}$ are complex electric and magnetic field amplitudes) corresponds to the projection onto the transversal subspace (where Fourier components of the fields are orthogonal to the wave vectors). 
The drawback of this approach to the Dirac equation is that it allows one to deal with only purely-electron (or positron) states, excluding the \emph{zitterbewegung} effects in mixed states. 
But in all cases where the type of particles is fixed, and the interband transitions (e.g., via scattering on external potentials) are negligible, this formalism perfectly describes the observable dynamics.

The covariant (projected) operators, defined via the Berry connection and curvature, are now routinely used in a variety of wave systems. 
Moreover, in the relativistic-electron context, such operators were introduced long before the discovery of the Berry phase \cite{61}. 
First of all, in 1948 Pryce published a comprehensive study \cite{30} of various possible position and AM operators for relativistic particles. 
There, his ``case (c)'' with the position operator ``$\mathbf{q}$'' and the corresponding spin ``$\mathbf{S}$'' exactly correspond to the projected operators $\boldsymbol{\mathcal{R}}$ and $\boldsymbol{\mathcal{S}}$ considered here. 
In terms of the classical many-particle analogue of a quantum distributed wavefunction, Pryce introduced this position as follows: ``{\it the coordinates of the mass-centre in a particular frame of reference is defined as the mean of the co-ordinates of the several particles weighted with their dynamical masses (energies)}''. 
One might think that this corresponds to the {\it center of energy} of the electron state. 
However this is not the case. 
Pryce calculated his operator using the center-of-energy operator $\mathbf{N} = \tfrac{1}{2}(\mathbf{r}H + H\mathbf{r})$ as $\mathbf{q} = \tfrac{1}{2}(H^{-1} {\bf N} + {\bf N} H^{-1})$. 
In fact, $\mathbf{q} = \boldsymbol{\mathcal{R}}$, and its expectation value for a single-electron state corresponds to the {\it center of the probability density} (center of charge), while the center of energy is defined as $\mathbf{r}_E = \langle {\bf N} \rangle / \langle H \rangle \neq \langle \mathbf{q} \rangle$. 
The difference is important, e.g., for the ``relativistic Hall effect'' (\ref{eq:11}), where the center of energy $\mathbf{r}_E$ undergoes the transverse shift twice as large as the center of the probability density \cite{54,55,57}. 
Second, the projected position operator $\boldsymbol{\mathcal{R}}$ and the spin-Hall effect corresponding to it appeared in 1959 in the work of Adams and Blount \cite{33} (up to some arithmetic inaccuracies therein). 
There, the Dirac-equation calculations are given in Appendix A as an example of application of the generic formalism describing electrons in solids. 
This approach anticipated the modern Berry-phase formalism \cite{39,40,41,43,44,45}. 
Third, a detailed analysis of the projected position operators for electrons and photons was provided in the papers \cite{34,35} by Fleming in 1965, in the book \cite{36} by Bacry in 1988, and also in a comprehensive monograph \cite{37} by Thaller in 1992. 
The connection of the projected spin operator with the Pauli-Lubanski vector was revealed by Czachor in 1997 \cite{38}. 
Finally, accurate Berry-phase descriptions with analyses of observable SOI effects in free space and in external fields was given by Berard and Mohrbach (2006) \cite{39}, Bliokh (2005) \cite{40}, Chang and Niu (2008) \cite{41}, and Bliokh {\it et al.}~(2011) \cite{9}.

The main equations and properties of the projected electron operators are summarized in Table \ref{table}.

\begin{table*}[t]
  \centering
  \begin{tabular}{|p{2.2cm}|p{6.5cm}|p{6.5cm}|}
\hline
\hline
& & \\
      &

{\bf Projected operators} 

&  

{\bf NWFW operators} 

\\
    \hline 

& & \\
    
    {\bf Standard 

representation}

& 
    
    $ \boldsymbol{\mathcal{R}}=\mathbf{r} + \dfrac{\mathbf{p}\times\mathbf{S}}{E^2}+ i \dfrac{m \beta \boldsymbol{\alpha}}{2E^2}$
    
    \emph{Pryce 1948} \cite{30}: first Eq.~(6.6); 
    
    \emph{Bacry 1988} \cite{36}: Eq.~(6.8), expressed via $\tilde{\mathbf{r}}$, $\tilde{\mathbf{S}}$;
    
    \emph{Thaller 1992} \cite{37}: Eq.~(1.145).
     
    & $\tilde{\mathbf{r}} = \mathbf{r} + \dfrac{\mathbf{p}\times\mathbf{S}}{E(E+m)}+ i \dfrac{\beta \boldsymbol{\alpha}}{2E}- i \dfrac{\beta(\boldsymbol{\alpha}\cdot\mathbf{p})\mathbf{p}}{2 E^2(E+m)}$ 
    
    \emph{Pryce 1948} \cite{30}: third Eq.~(6.6); 
    
    \emph{Newton, Wigner 1949} \cite{31}: Eq.~(2.2); 

    \emph{Foldy, Wouthuysen 1950} \cite{32}: Eq.~(23) 

[with an arithmetic inaccuracy]; 
        
    \emph{Thaller 1992} \cite{37}: Eq.~(1.158).

    \\
    
    & $\boldsymbol{\mathcal{S}} = \dfrac{m^2}{E^2} \mathbf{S} + \dfrac{(\mathbf{p}\cdot\mathbf{S})\mathbf{p}}{E^2} - i \dfrac{m \beta (\boldsymbol{\alpha} \times \mathbf{p})}{2E^2}$
    
    \emph{Pryce 1948} \cite{30}: first Eq.~(6.7); 
    
    \emph{Thaller 1992} \cite{37}: Eq.~(1.151);

    \emph{Czachor 1997} \cite{38}: Eq.~(23). 
    
    & $\tilde{\mathbf{S}} = \dfrac{m}{E} \mathbf{S} + \dfrac{(\mathbf{p}\cdot\mathbf{S})\mathbf{p}}{E(E+m)} - i \dfrac{\beta (\boldsymbol{\alpha} \times \mathbf{p})}{2E}$

    \emph{Pryce 1948} \cite{30}: third Eq.~(6.7); 
    
    \emph{Foldy, Wouthuysen 1950} \cite{32}: Table I 

[global $1/2$ factor is missing];
    
    \emph{Barnett 2017} \cite{18}: Eq.~(1.11) 

[$1/2$ missing in the second term].
    
    \\
    
    \hline & & \\ {\bf FW 

representation}

&
    
    $\boldsymbol{\mathcal{R}}_{\mathrm{FW}} = \mathbf{r} + \dfrac{\mathbf{p}\times \mathbf{S}}{E(E+m)}$ 
    
    \emph{Adams, Blount 1959} \cite{33}: the first line of Eq.~(A1b) [the second term has incorrect sign];
    
    \emph{Berard, Mohrbach 2006} \cite{39}: Eq.~(10);
    
    \emph{Bliokh 2005} \cite{40}: Eqs.~(8) and (10).
    
    &
    
    $\tilde{\mathbf{r}}_{\mathrm{FW}} = \mathbf{r}$
    
    \emph{Pryce 1948} \cite{30}: third Eq.~(6.9);
    
    \emph{Foldy, Wouthuysen 1950} \cite{32}: Table I;
    
    \emph{Thaller 1992} \cite{37}: Eq.~(1.168). \\

    & $\boldsymbol{\mathcal{S}}_{\mathrm{FW}} = \dfrac{m}{E}\mathbf{S} + \dfrac{(\mathbf{p}\cdot\mathbf{S})\mathbf{p}}{E(E+m)}$;
    
    \emph{Bliokh, Dennis, Nori 2011} \cite{9}: Eq.~(15). 
    
    & $\tilde{\mathbf{S}}_{\mathrm{FW}} = \mathbf{S}$
    
    \emph{Pryce 1948} \cite{30}: third Eq. (6.9); 
    
    \emph{Foldy, Wouthuysen 1950} \cite{32}: Table I 

[global 1/2 factor is missing]; 
    
    \emph{Barnett 2017} \cite{18}. 

\\
  
    \hline   

& & \\

{\bf Properties} 
    
    & Proper time evolution; 
    
    Canonical expectation values; 
    
    Non-canonical commutators; 
    
    Smooth massless limit; 
    
    Naturally describe SOI and Berry-phase phenomena; 
    
    Spin is the spatial part of the relativistic Pauli-Lubanski vector.
    
    & Proper time evolution; 
    
    Non-canonical expectation values;
    
    Canonical commutators;
    
    Singular massless limit; 
    
    Lack SOI and Berry-phase phenomena;
    
    $ $
     
    Spin is the non-relativistic rest-frame spin.
    \\ \hline
\hline
  \end{tabular}
  \caption{Explicit expressions for the ``projected'' and NWFW operators in the standard and FW representations. 
Key works deriving and analyzing these expressions are listed. 
In all cases, the orbital AM operator is given by the vector product of the corresponding positions operator and momentum $\mathbf{p}$. 
The last row lists the main physical features of the two sets of operators.}\label{table}
\end{table*}

\subsection{``Newton-Wigner-Foldy-Wouthuysen'' operators}\label{sub:4.2}

An alternative way to construct relativistic-electron operators with proper time evolution is to use the {\it inverse FW transformation} instead of the projection (\ref{eq:12}) \cite{32}:
\begin{align}
   \tilde{\mathbf{r}} = U_{\mathrm{FW}}^{\dagger} \mathbf{r} U_{\mathrm{FW}}, \nonumber \\
\tilde{\mathbf{L}} = U_{\mathrm{FW}}^{\dagger} \mathbf{L} U_{\mathrm{FW}} = \tilde{\mathbf{r}} \times \mathbf{p}, ~~
   \tilde{\mathbf{S}} & = U_{\mathrm{FW}}^{\dagger} \mathbf{S} U_{\mathrm{FW}} = \mathbf{J} - \tilde{\mathbf{L}}. 
   \label{eq:21}
\end{align}
Obviously, in the FW representation these operators acquire the canonical forms
\begin{equation}
   \tilde{\mathbf{r}}_{\mathrm{FW}} = \mathbf{r}, \quad \tilde{\mathbf{L}}_{\mathrm{FW}} = \mathbf{L} = \mathbf{r}\times\mathbf{p}, \quad \tilde{\mathbf{S}}_{\mathrm{FW}} = \mathbf{S}. 
   \label{eq:22}
\end{equation}
From here, using $H_{\mathrm{FW}} = \beta E$, one can readily see that these operators obey the proper time evolution similar to Eqs.~(\ref{eq:17}),
\begin{equation}
   \frac{d\tilde{\mathbf{r}}}{dt} = i[H,\tilde{\mathbf{r}}] = \mathbf{p}H^{-1}, \quad [H,\tilde{\mathbf{S}}] = [H,\tilde{\mathbf{L}}] =0. 
   \label{eq:23}
\end{equation}
One can also see that these operators obey canonical commutation relations [cf.~Eqs.~(\ref{eq:19}) and (\ref{eq:20})]
\begin{equation}
   [\tilde{r}_i,\tilde{r}_j] = 0,\quad
   [\tilde{S}_i,\tilde{S}_j] = i \varepsilon_{ijk} \tilde{S}_k,\quad
   [\tilde{L}_i,\tilde{L}_j] = i \varepsilon_{ijk} \tilde{L}_k.
   \label{eq:24}
\end{equation}  
In the standard representation, the position and spin operators (\ref{eq:21}) read:
\begin{align}
   \tilde{\mathbf{r}} & = \mathbf{r} + \frac{\mathbf{p}\times \mathbf{S}}{E(E+m)} + 
i \frac{\beta\boldsymbol{\alpha}}{2E} -
i \frac{\beta (\boldsymbol{\alpha}\cdot \mathbf{p})\mathbf{p}}{2E^2 (E+m)}, \nonumber\\ 
   \tilde{\mathbf{S}} & = \frac{m}{E}\mathbf{S} + \frac{(\mathbf{p}\cdot \mathbf{S})\mathbf{p}}{E(E+m)} - i \frac{\beta(\boldsymbol{\alpha}\times \mathbf{p})}{2E}.
   \label{eq:25}
\end{align}
   
The operator $\tilde{\mathbf{r}}$ is well known as the Newton-Wigner position operator \cite{31,32,34,35,36,37}, while the general approach (\ref{eq:21})--(\ref{eq:25}) was described by Foldy and Wouthuysen \cite{32}. 
Therefore we refer to the operators (\ref{eq:21})--(\ref{eq:25}) as the {\it Newton-Wigner-Foldy-Wouthuysen (NWFW)} operators. 
Despite the proper time evolution (\ref{eq:23}), these operators have a serious drawback, namely, their {\it expectation values differ from the canonical ones} [cf.~Eqs.~(\ref{eq:18})]:
\begin{equation}
   \langle \tilde{\mathbf{r}} \rangle \neq \langle \mathbf{r} \rangle, \quad \langle \tilde{\mathbf{S}} \rangle \neq \langle \mathbf{S} \rangle, \quad \langle \tilde{\mathbf{L}} \rangle \neq \langle \mathbf{L} \rangle,
   \label{eq:26}
\end{equation}
For example, using the FW representation $\tilde{\mathbf{S}}_{\mathrm{FW}} = \mathbf{S}$, we easily see that the expectation value of the
NWFW spin for a plane electron wave (\ref{eq:2}) yields
\begin{equation}
   \langle \tilde{\mathbf{S}} \rangle = \langle \mathbf{s} \rangle.
   \label{eq:27}
\end{equation}
Evidently, this is the \emph{non-relativistic spin in the electron rest frame} instead of the relativistic momentum-dependent spin (\ref{eq:7}). 
Similarly, the spin-to-orbital AM conversion (\ref{eq:10}) in nonparaxial electron vortex beams is missing for the NWFW operators:
\begin{equation}
   \langle \tilde{S}_z \rangle = \langle s_z \rangle, \quad \langle \tilde{L}_z \rangle = \ell.
   \label{eq:28}
\end{equation}
Thus, \emph{relativistic transformations of the dynamical properties of the electron and the SOI phenomena are missing for these operators}. 
This contradicts numerous observable SOI effects (spin-dependent orbital characteristics), known for both electron \cite{9,39,40,41,43,44,45,46,62} and optical (photon) \cite{24,25,47,48,49,50,51,52,53} fields.

Note that the most standard textbook example of the SOI energy responsible for the fine structure of atomic levels is naturally derived from the projected coordinate operator $\boldsymbol{\mathcal{R}}$ using the non-relativistic limit ($p \ll m$) in the FW representation, where the electron wavefunction is two-component \cite{35,36,62}. 
Considering a spherically-symmetric potential $V(r)$, we obtain, in the FW representation
\begin{align}
   \boldsymbol{\mathcal{R}}_{\mathrm{FW}}^2 & \simeq \mathbf{r}^2 + \frac{\mathbf{r}\cdot(\mathbf{p} \times \mathbf{S})}{m^2} = \mathbf{r}^2 + \frac{\mathbf{L}\cdot\mathbf{S}}{m^2}, \nonumber \\
   V(|\boldsymbol{\mathcal{R}}_{\mathrm{FW}}|) & \simeq V(r) + \frac{d V(r)}{dr} \frac{\mathbf{L}\cdot\mathbf{S}}{2 m^2 r}. 
   \label{eq:29}
\end{align}
Here the correction term (with the trivial reduction $\mathbf{S} \to \mathbf{s}$) is the well-known SOI energy \cite{46}. 
Equation (\ref{eq:29}) shows that the NWFW position operator $\tilde{\mathbf{r}}_{\mathrm{FW}} = \mathbf{r}$ corresponds to the {\it canonical}
coordinates in the Pauli Hamiltonian with a separate SOI term (absent in the full Dirac equation!), while the projected position $\boldsymbol{\mathcal{R}}_{\mathrm{FW}}$ describes the {\it covariant} coordinates, and the SOI is \emph{intrinsically} present in the potential-energy term $V(r)$. 
Importantly, it is the covariant (i.e., projected) coordinates that correspond to the actual centroid of a localized electron state. 
Moreover, the covariant coordinates determine the equations of motions of the electron in smooth external potentials, while canonical coordinates can produce erroneous results, as shown in \cite{39,63,64}.

Another important drawback of the NWFW operators is that they \emph{cannot be extended to the case of massless particles} (e.g., Weyl particles or photons) \cite{31,32,35,36}. 
Therefore, this approach is \emph{singular} in the $m \to 0$ limit. 
This is explained by the fact that such operators are associated with the rest-frame properties of the electron, while there is no rest frame for massless particles. 
This feature makes the NWFW approach not suitable for condensed matter systems. 
There, effective masses (gaps in the spectra) can vary, passing via zeros (Dirac or Weyl points), depending on tunable parameters of the system. 
Therefore, the description of solid-state electrons requires a formalism depending smoothly on $m$. 
Moreover, solid-state electrons exhibit numerous observable SOI effects, which underlie the field of spintronics.

Historically, the NWFW operators first appeared in 1948 in the same work \cite{30} by Pryce. 
They correspond to the ``case (e)'' and position ``$\tilde{\mathbf{q}}$'' and spin ``$\tilde{\mathbf{S}}$''. 
Pryce derived these operators to ``improve'' the non-canonical commutation relations (\ref{eq:19}) and (\ref{eq:20}). 
He also explicitly mentioned the close relation of the position operator $\tilde{\mathbf{q}} = \tilde{\mathbf{r}}$ to the rest frame and impossibility to use it for photons. 
One year later, Newton and Wigner suggested the operator $\tilde{\mathbf{r}}$ again \cite{31} using arguments related to the localizability of massive quantum particles. 
Finally, the whole approach, including spin and orbital AM $\tilde{\mathbf{S}}$ and $\tilde{\mathbf{L}}$, was described in 1950 by Foldy and Wouthuysen \cite{32} (up to an arithmetic inaccuracy in $\tilde{\mathbf{r}}$). 
This approach was criticized by Bacry \cite{36} in favor of the projected-operators formalism. 
Finally, very recently Barnett revisited this formalism in \cite{18} suggesting the same spin and orbital AM $\tilde{\mathbf{S}}$ and $\tilde{\mathbf{L}}$ (up to a missing factor of $1/2$ in the $\beta(\boldsymbol{\alpha}\times \mathbf{p})$-terms), which are free of the SOI.

The main equations and properties of the NWFW electron operators are summarized in Table I.

\section{Discussion}\label{sec:5}

In conclusion, we have reviewed and compared two approaches to the description of the position, spin, and angular momentum (AM) of a relativistic electron. 
The first one is based on the projection of canonical Dirac operators onto the positive-energy (electron) subspace, whereas the second one assumes canonical form of operators in the Foldy-Wouthuysen (FW) representation. 
We have shown that the ``projected'' formalism results in the same observable phenomena and expectation values as the canonical Dirac approach, while elucidating the spin-orbit interaction (SOI) effects via the Berry-phase formalism. 
In turn, the second formalism produces the ``Newton-Wigner-Foldy-Wouthuysen'' (NWFW) operators with essentially different physical properties. 
Most importantly, because of the close relation of the NWFW operators to the rest-frame properties of the electron, this approach lacks SOI phenomena (in the full Dirac treatment) and has a singular zero-mass limit.

The following qualitative arguments could shed some light on the peculiarities of the NWFW operators. 
First, the FW transformation (\ref{eq:3}) is defined in the momentum representation, and it transforms the electron plane-wave bispinor (\ref{eq:2}) to the spinor $W_{\mathrm{FW}} =(w,0)^T$. 
Consider now the \emph{Lorentz boost} of an electron plane wave to the rest frame. It is given by the non-unitary Hermitian operator $\Lambda = \frac{E+m-\boldsymbol{\alpha}\cdot \mathbf{p}}{\sqrt{2E(E+m)}},$ resembling $U_{\mathrm{FW}}$ and transforming the bispinor (\ref{eq:2}) to $W' = \Lambda W = \sqrt{m/E} (w,0)^T$.
Thus, one can regard the FW transformation as a ``unitary counterpart of the Lorentz boost to the rest frame''.
This explains why the NWFW operators, chosen as ``canonical in the FW representation'' describe some rest-frame properties of the electron, such as the rest-frame spin (\ref{eq:27}).

Second, we consider the transition from the Dirac equation to the non-relativistic Schr\"odinger equation with Pauli Hamiltonian (including the SOI term) and two-component wavefunction $\varphi_{\mathrm{Pauli}}$.
Writing the Dirac bispinor wavefunction as $\psi = (\varphi,\chi)^T,$ the Pauli wavefunction is given by \cite{46} $\varphi_{\mathrm{Pauli}} \simeq \left(1+\frac{p^2}{8m^2}\right)\varphi$, where we used $p \ll m$ and omitted the phase factor $\exp(imt/\hbar)$. 
In the same approximation, using $\chi \simeq \frac{\boldsymbol{\sigma}\cdot\mathbf{p}}{2m}\varphi$ and $U_{\mathrm{FW}} \simeq 1 + \frac{\beta \boldsymbol{\alpha}\cdot\mathbf{p}}{2m} - \frac{p^2}{8m^2}$, the FW wavefunction $\psi_{\mathrm{FW}} = (\varphi_{\mathrm{FW}},0)$ reduces to the \emph{same Pauli spinor}: $\varphi_{\mathrm{FW}} \simeq \varphi_{\mathrm{Pauli}}$. 
Thus, one can say that the non-relativistic two-component Pauli wavefunction corresponds to the relativistic \emph{FW wavefunction}. 
Therefore, the NWFW position operator $\tilde{\mathbf{r}}$, having canonical form in the FW representation, appears as the \emph{canonical} coordinates $\mathbf{r}$ in the Pauli Hamiltonian, and an additional SOI term arises there. 
However, one should remember that the FW transformation is {\it nonlocal} in real space, and this nonlocality is hidden in the Pauli Hamiltonian and wavefunction. 
Using the canonical Dirac position and projecting it onto the electron subspace results in the position operator $\boldsymbol{\mathcal{R}}_{\mathrm{FW}}$ (with the trivial reduction $\mathbf{S} \to \mathbf{s}$) in the Pauli formalism. 
This operator corresponds to \emph{covariant} coordinates, which describe the electron centroid and determine the covariant equations of motion \cite{33,39,40,41,43,44,45,62,63,64}. 
Moreover, this position operator reveals the intrinsic nonlocality of the Pauli formalism via anomalous commutation relations (\ref{eq:19}) and unveils the geometric Berry-phase origin of the SOI term (\ref{eq:29}) in the Pauli Hamiltonian \cite{39,40,41,43,44,45,62}. 
In terms of covariant operators, the SOI does not require additional terms in the Hamiltonian but appears as an inherent electron feature, as in the full Dirac equation.

In this work we mostly considered properties of the Dirac electron in {\it free space}, i.e., without external potentials. 
In the presence of potentials, the problem is complicated considerably. 
Indeed, in this case the notion of a pure {\it electron} does not make sense, and the combined {\it electron-positron} description becomes necessary.
Therefore, rigorously speaking, the projected operators are applicable in external potentials only in the adiabatic (semiclassical) approximation, when the electron-positron transitions are negligible. 
Nonetheless, even in scattering problems with electron-positron transitions, the incoming and outgoing states of relativistic electrons and positrons can be described using projected operators (e.g., the Pauli-Lubanski spin vector). 
A detailed comparison of various spin definitions in the presence of external potentials was recently provided in \cite{42}.

We finally note that, rigorously speaking, the domains of applicability of the operators under discussion imply square-integrable electron wavefunctions localized in three spatial dimensions. 
In this manner, all of the explicit examples mentioned in \cite{9,17,18} and in the present work should be considered as simplified illustrations, while a more accurate wavepacket treatment may reveal additional fine features.

We acknowledge helpful discussions with Michael Berry, Iwo Bialynicki-Birula, Michael Stone, and Heiko Bauke. 
This work was supported by the RIKEN iTHES Project, MURI Center for Dynamic Magneto-Optics via the AFOSR Award No.~FA9550-14-1-0040, the Japan Society for the Promotion of Science (KAKENHI), the IMPACT program of JST, CREST grant No.~JPMJCR1676, the John Templeton Foundation, and the Australian Research Council.


\begin{thebibliography}{10}

\bibitem{1} L.~Allen, S.M.~Barnett, and M.J.~Padgett, \emph{Optical Angular Momentum} (IoP Publishing, Bristol, 2003).
\bibitem{2} D.L.~Andrews and M.~Babiker, \emph{The Angular Momentum of Light} (Cambridge University Press, Cambridge, 2013)
\bibitem{3} K.Y.~Bliokh, Y.P.~Bliokh, S.~Savel'ev, and F.~Nori, Semiclassical dynamics of electron wave packet states with phase vortices, Phys. Rev. Lett. {\bf 99}, 190404 (2007).
\bibitem{4} M.~Uchida and A.~Tonomura, Generation of electron beams carrying orbital angular momentum, Nature {\bf 464}, 737 (2010).
\bibitem{5} J.~Verbeeck, H.~Tian, and P.~Schattschneider, Production and application of electron vortex beams, Nature {\bf 467}, 301 (2010).
\bibitem{6} B.J.~McMorran, A.~Agrawal, I.M.~Anderson, A.A.~Herzing, H.J.~Lezec, J.J.~McClelland, and J.~Unguris, Electron vortex beams with high quanta of orbital angular momentum, Science {\bf 331}, 192 (2011).
\bibitem{7} J.~Harris, V.~Grillo, E.~Mafakheri, G.C.~Gazzadi, S.~Frabboni, R.W.~Boyd, and E.~Karimi, Structured quantum waves, Nature Phys. {\bf 11}, 629 (2015).
\bibitem{8} K.Y.~Bliokh, I.P.~Ivanov, G.~Guzzinati, L.~Clark, R.~Van Boxem, A.~B{\'e}ch{\'e}, R.~Juchtmans, M.A.~Alonso, P.~Schattschneider, F.~Nori, and J.~Verbeeck, Theory and applications of free-electron vortex states, Phys. Rep. {\bf 690}, 1 (2017).
\bibitem{9} K.Y.~Bliokh, M.R.~Dennis, and F.~Nori, Relativistic electron vortex beams: Angular momentum and spin-orbit interaction, Phys. Rev. Lett. {\bf 107}, 174802 (2011).
\bibitem{10} D.V.~Karlovets, Electron with orbital angular momentum in a strong laser wave, Phys. Rev. A {\bf 86}, 62102 (2012).
\bibitem{11} A.G.~Hayrapetyan, O.~Matula, A.~Aiello, A.~Surzhykov, and S.~Fritzsche, Interaction of relativistic electron-vortex beams with few-cycle laser pulses, Phys. Rev. Lett. {\bf 112}, 134801 (2014).
\bibitem{12} V.~Serbo, I.P.~Ivanov, S.~Fritzsche, D.~Seipt, and A.~Surzhykov, Scattering of twisted relativistic electrons by atoms, Phys. Rev. A {\bf 92}, 012705 (2015).
\bibitem{13} I.P.~Ivanov, D.~Seipt, A.~Surzhykov, and S.~Fritzsche, Elastic scattering of vortex electrons provides direct access to the Coulomb phase, Phys. Rev. D {\bf 94}, 076001 (2016).
\bibitem{14} I.~Kaminer {\it et al.}, Quantum {\v C}erenkov radiation: Spectral cutoffs and the role of spin and orbital angular momentum, Phys. Rev. X {\bf 6}, 011006 (2016).
\bibitem{15} I.P.~Ivanov, V.G.~Serbo, and V.A.~Zaytsev, Quantum calculation of the Vavilov-Cherenkov radiation by twisted electrons, Phys. Rev. A {\bf 93}, 053825 (2016).
\bibitem{16} R.V.~Boxem, J.~Verbeeck, and B.~Partoens, Spin effects in electron vortex states, EPL {\bf 102}, 40010 (2013).
\bibitem{17} I.~Bialynicki-Birula and Z.~Bialynicka-Birula, Relativistic electron wave packets carrying angular momentum, Phys. Rev. Lett. {\bf 118}, 114801 (2017).
\bibitem{18} S.M.~Barnett, Relativistic electron vortices, Phys. Rev. Lett. {\bf 118}, 114802 (2017).
\bibitem{19} P.~Schattschneider and J.~Verbeeck, Theory of free electron vortices, Ultramicroscopy {\bf 111}, 1461 (2011).
\bibitem{20} M.R.~Dennis, K.~O'Holleran, and M.J.~Padgett, Singular optics: Optical vortices and polarization singularities, Prog. Opt. {\bf 53}, 293 (2009).
\bibitem{21} K.Y.~Bliokh and F.~Nori, Transverse and longitudinal angular momenta of light, Phys. Rep. {\bf 592}, 1 (2015).
\bibitem{22} K.Y.~Bliokh, A.Y.~Bekshaev, and F.~Nori, Extraordinary momentum and spin in evanescent waves, Nature Commun. {\bf 5}, 3300 (2014).
\bibitem{23} M.~Antognozzi {\it et al}., Direct measurements of the extraordinary optical momentum and transverse spin-dependent force using a nano-cantilever, Nature Phys. {\bf 12}, 731 (2016).
\bibitem{24} K.Y.~Bliokh, M.A.~Alonso, E.A.~Ostrovskaya, and A.~Aiello, Angular momenta and spin-orbit interaction of nonparaxial light in free space, Phys. Rev. A {\bf 82}, 063825 (2010).
\bibitem{25} K.Y.~Bliokh, F.J.~Rodr{\'i}guez-Fortu{\~n}o, F.~Nori, and A.V.~Zayats, Spin-orbit interactions of light, Nature Photon. {\bf 9}, 796 (2015).
\bibitem{26} P.~Ouyang, V.~Mohta, and R.L.~Jaffe, Dirac particles in twisted tubes, Ann. Phys. {\bf 275}, 297 (1999).
\bibitem{27} C.C.~Leary, D.~Reeb, and M.G.~Raymer, Self-spin-controlled rotation of spatial states of a Dirac electron in a cylindrical potential via spin-orbit interaction, New J. Phys. {\bf 10}, 103022 (2008).
\bibitem{28} K.~Huang, On the zitterbewegung of the Dirac electron, Am. J. Phys. {\bf 20}, 479 (1952).
\bibitem{29} A.O.~Barut and A.J.~Bracken, Magnetic-moment operator of the relativistic electron, Phys. Rev. D {\bf 24}, 3333 (1981).
\bibitem{30} M.H.L.~Pryce, The mass-centre in the restricted theory of relativity and its connexion with the quantum theory of elementary particles, Proc. R. Soc. A {\bf 195}, 62 (1948).
\bibitem{31} T.D.~Newton and E.P.~Wigner, Localized states for elementary systems, Rev. Mod. Phys. {\bf 21}, 400 (1949).
\bibitem{32} L.L.~Foldy and S.A.~Wouthuysen, On the Dirac theory of spin 1/2 particles and its non-relativistic limit, Phys. Rev. {\bf 78}, 29 (1950).
\bibitem{33} E.N.~Adams and E.I.~Blount, Energy bands in the presence of an external force field -- II. Anomalous velocities, J. Phys. Chem. Solids {\bf 10}, 286 (1959).
\bibitem{34} G.N.~Fleming, Covariant position operators, spin and locality, Phys. Rev. {\bf 137}, B188 (1965).
\bibitem{35} G.N.~Fleming, Nonlocal properties of stable particles, Phys. Rev. {\bf 139}, B963 (1965).
\bibitem{36} H.~Bacry, \emph{Localizability and Space in Quantum Physics} (Springer, Berlin, 1988).
\bibitem{37} B.~Thaller, \emph{The Dirac Equation} (Springer, Berlin, 1992).
\bibitem{38} M.~Czachor, Einstein-Podolsky-Rosen-Bohm experiment with relativistic massive particles, Phys. Rev. A {\bf 55}, 72 (1997).
\bibitem{39} A.~B{\'e}rard and H. Mohrbach, Spin Hall effect and Berry phase of spinning particles, Phys. Lett. A {\bf 352}, 190 (2006).
\bibitem{40} K.Y.~Bliokh, Topological spin transport of a relativistic electron, Europhys. Lett. {\bf 72}, 7 (2005).
\bibitem{41} M.-C.~Chang and Q.~Niu, Berry curvature, orbital moment and effective quantum theory of electrons in electromagnetic fields, J. Phys.: Condens. Matter {\bf 20}, 193202 (2008).
\bibitem{42} H.~Bauke, S.~Ahrens, C.H.~Keitel, and R.~Grobe, Relativistic spin operators in various electromagnetic environments, Phys. Rev. A {\bf 89}, 052101 (2014).
\bibitem{43} P.~Gosselin, A.~B{\'e}rard, and H.~Mohrbach, Semiclassical dynamics of Dirac particles interacting with a static gravitational field, Phys. Lett. A {\bf 368}, 356 (2007).
\bibitem{44} P.~Gosselin and H.~Mohrbach, Diagonal representation for a generic matrix valued quantum Hamiltonian, Eur. Phys. J. C {\bf 64}, 495 (2009).
\bibitem{45} D.~Xiao, M.-C.~Chang, and Q.~Niu, Berry phase effects on electronic properties, Rev. Mod. Phys. {\bf 82}, 1959 (2010).
\bibitem{46} V.B.~Berestetskii, E.M.~Lifshitz, and L.P.~Pitaevskii, \emph{Quantum Electrodynamics} (Pergamon, Oxford, 1982).
\bibitem{47} A.~Dogariu and C.~Schwartz, Conservation of angular momentum of light in single scattering, Opt. Express {\bf 14}, 8425 (2006).
\bibitem{48} Y.~Zhao, J.S.~Edgar, G.D.M.~Jeffries, D.~McGloin, and D.T.~Chiu, Spin-to-orbital angular momentum conversion in a strongly focused optical beam, Phys. Rev. Lett. {\bf 99}, 073901 (2007).
\bibitem{49} T.A.~Nieminen, A.B.~Stilgoe, N.R.~Heckenberg, and H.~Rubinsztein-Dunlop, Angular momentum of a strongly focused Gaussian beam, J. Opt. A {\bf 10}, 115005 (2008).
\bibitem{50} K.Y.~Bliokh {\it et al.}, Spin-to-orbit angular momentum conversion in focusing, scattering, and imaging systems. Opt. Express {\bf 19}, 26132 (2011).
\bibitem{51} K.Y.~Bliokh and Y.P.~Bliokh, Topological spin transport of photons: The optical Magnus effect and Berry phase, Phys. Lett. A {\bf 333}, 181 (2004).
\bibitem{52} M.~Onoda, S.~Murakami, and N.~Nagaosa, Hall effect of light, Phys. Rev. Lett. {\bf 93}, 083901 (2004).
\bibitem{53} K.Y.~Bliokh, A.~Niv, V.~Kleiner, and E.~Hasman, Geometrodynamics of spinning light, Nature Photon. {\bf 2}, 748 (2008).
\bibitem{54} R.A.~Muller, Thomas precession: Where is the torque?, Am. J. Phys. {\bf 60}, 313 (1992).
\bibitem{55} K.Y.~Bliokh and F.~Nori, Relativistic Hall effect, Phys. Rev. Lett. {\bf 108}, 120403 (2012).
\bibitem{56} J-Y.~Chen, D.T.~Son, M.A.~Stephanov, H-U.~Yee, and Y.~Yin, Lorentz invariance in chiral kinetic theory, Phys. Rev. Lett. {\bf 113}, 182302 (2014).
\bibitem{57} M.~Stone, V.~Dwivedi, and T.~Zhou, Wigner translations and the observer dependence of the position of massless spinning particles, Phys. Rev. Lett. {\bf 114}, 210402 (2015).
\bibitem{58} S.J.~Van Enk and G.~Nienhuis, Commutation rules and eigenvalues of spin and orbital angular momentum of radiation fields, J. Mod. Opt. {\bf 41}, 963 (1994).
\bibitem{59} I.~Bialynicki-Birula and Z.~Bialynicka-Birula, Canonical separation of angular momentum of light into its orbital and spin parts, J. Opt. {\bf 13}, 064014 (2011).
\bibitem{60} B.S.~Skagerstam, Localization of massless spinning particles and the Berry phase, arXiv:hep-th/9210054 (1992).
\bibitem{61} M.V.~Berry, Quantal phase factors accompanying adiabatic changes, Proc. R. Soc. Lond. A {\bf 392}, 45 (1984).
\bibitem{62} H.~Mathur, Thomas precession, spin-orbit interaction, and Berry's phase, Phys. Rev. Lett. {\bf 67}, 3325 (1991).
\bibitem{63} S.-Q.~Shen, Spin transverse force on spin current in an electric field, Phys. Rev. Lett. {\bf 95}, 187203 (2005).
\bibitem{64} K.Y.~Bliokh, Comment on ``Spin transverse force on spin current in an electric field'', arXiv:cond-mat/0511146 (2005).

\end{thebibliography}
\end{document}